\documentclass[letterpaper, 10pt, conference]{ieeeconf}
\IEEEoverridecommandlockouts

\usepackage{amsmath,amssymb,amsthm,mathtools,bm}
\usepackage{cite}
\usepackage{microtype}
\usepackage{pgfplots}
\usepackage{circuitikz}
\usepackage{booktabs}
\usepackage{url}
\usepackage{enumerate}
\usepackage{algorithm}
\usepackage{algpseudocode}
\pgfplotsset{compat=1.18}
\ctikzset{bipoles/length=0.8cm}
\usepgfplotslibrary{groupplots}
\newcommand\NMpanelwidth{5.2cm}
\newcommand\NMpanelheight{3.3cm}

\definecolor{nmSpice}{HTML}{EB6834}
\definecolor{nmZeroBias}{HTML}{2A78D6}
\definecolor{nmSector}{HTML}{1BAF7A}
\definecolor{nmInk}{HTML}{52514E}
\definecolor{nmGrid}{HTML}{E6E5E0}

\pgfplotsset{
  table/nm csv/.style={col sep=comma, x=N},
  nm series/.style={color=#1, line width=0.9pt, mark options={fill=#1, draw=white, line width=0.4pt, solid}},
  nm spice/.style={nm series=nmSpice, mark=*, mark size=1.8pt},
  nm zero bias/.style={nm series=nmZeroBias, mark=square*, mark size=1.6pt},
  nm sector/.style={nm series=nmSector, mark=triangle*, mark size=2.2pt},
  nm bars/.style={error bars/.cd, y dir=both, y explicit, error bar style={line width=0.6pt, solid},
    error mark options={rotate=90, mark size=1.6pt, line width=0.6pt}},
  table/nm range/.style={y error plus expr=\thisrow{#1_max}-\thisrow{#1},
    y error minus expr=\thisrow{#1}-\thisrow{#1_min}},
  nm timeout/.style={only marks, mark=x, mark size=3.2pt, line width=1.3pt, color=#1, forget plot,
    nodes near coords={timed out}, every node near coord/.append style={font=\scriptsize, text=#1, anchor=south east,
    xshift=-1pt}},
  nm cap/.style={draw=nmInk!70, line width=0.6pt, dash pattern=on 3pt off 2pt, forget plot},
  nm reference/.style={draw=nmInk!55, line width=0.6pt, dashed, forget plot},
  nm axis/.style={
    width=\NMpanelwidth, height=\NMpanelheight, scale only axis,
    xmode=log, ymode=log,
    xtick={16,64,256,1024}, xticklabels={16,64,256,1024}, minor xtick={32,128,512},
    xmin=12, xmax=1400,
    yminorticks=false,
    xlabel={$N$ ($N\times N$ array)},
    grid=major, major grid style={draw=nmGrid, line width=0.4pt},
    axis lines=left, axis line style={draw=nmInk!60, -},
    tick style={draw=nmInk!60}, tick label style={font=\scriptsize, text=nmInk},
    label style={font=\footnotesize, text=nmInk}, title style={font=\small, text=black, yshift=-2pt},
    ylabel style={yshift=-2pt},
    unbounded coords=jump,
  },
  group/nm grid style/.style={group size=3 by #1, horizontal sep=0.3cm, vertical sep=0.3cm,
    x descriptions at=edge bottom, y descriptions at=edge left},
  nm runtime row/.style={ymin=1e-3, ymax=5e3, ytick={1e-3,1e-2,1e-1,1,1e1,1e2,1e3}},
  nm speedup row/.style={ymin=0.2, ymax=60, ytick={0.25,1,4,16}, log ticks with fixed point},
  nm iterations row/.style={ymin=2, ymax=1e4, ytick={1e1,1e2,1e3,1e4}},
  nm per iteration row/.style={ymin=0.05, ymax=5e4, ytick={1e-1,1,1e1,1e2,1e3,1e4}},
  nm memory row/.style={ymin=0.05, ymax=40, ytick={0.1,1,10}, log ticks with fixed point},
  nm legend/.style={legend columns=3, legend style={draw=none, font=\footnotesize, text=nmInk,
    /tikz/every even column/.append style={column sep=0.8em}}},
}

\theoremstyle{plain}
\newtheorem{theorem}{Theorem}
\newtheorem{lemma}[theorem]{Lemma}
\newtheorem{proposition}[theorem]{Proposition}

\theoremstyle{definition}

\theoremstyle{remark}
\newtheorem{remark}{Remark}
\newtheorem{example}{Example}

\newcommand{\R}{\mathbb{R}}
\newcommand{\Z}{\mathbb{Z}}
\newcommand{\set}[1]{\mathsf{#1}}
\newcommand{\B}{\set{B}}
\newcommand{\V}{\set{V}}
\newcommand{\I}{\set{I}}
\newcommand{\Cc}{\set{C}}

\newcommand{\E}{\mathbb{E}}
\newcommand{\Vertex}{\mathbb{V}}

\newcommand{\ran}{\operatorname{ran}}

\newcommand{\diag}{\operatorname{diag}}
\newcommand{\gra}{\operatorname{graph}}
\newcommand{\dom}{\operatorname{dom}}
\newcommand{\zer}{\operatorname{zer}}
\newcommand{\fix}{\operatorname{fix}}

\newcounter{algo}
\renewcommand{\thealgo}{\arabic{algo}}

\newcommand{\bb}{\mathbb}

\renewcommand{\ker}{\mathrm{ker}}

\usepackage[dvipsnames]{xcolor}
\definecolor{rcBlue}{RGB}{37,99,235}
\definecolor{rcPink}{RGB}{219,39,119}
\definecolor{rcGray}{RGB}{160,160,160}
\definecolor{rcTeal}{RGB}{13,148,136}

\title{\LARGE\bf Simulating Nonlinear Resistive Networks using a Behavioral Splitting}

\author{Ryan Alizadeh, Thomas Chaffey, Dominic Liao-McPherson, and Alberto Padoan\\
\thanks{R. Alizadeh and A. Padoan are with the Department of Electrical and Computer Engineering, University of British Columbia, Vancouver, BC V6T 1Z4, Canada (email: {\tt\small ryan.alizadeh@ubc.ca, alberto.padoan@ubc.ca}).
T. Chaffey is with the School of Electrical and Computer Engineering, The University of Sydney, Sydney, Australia (email: {\tt\small thomas.chaffey@sydney.edu.au}).
D. Liao-McPherson is with the Department of Mechanical Engineering, University of British Columbia, Vancouver, BC V6T 1Z4, Canada (email: {\tt\small dliaomcp@mech.ubc.ca}).
This work was supported by the Natural Sciences and Engineering Research Council of Canada under grants RGPIN-2025-06895 and DGECR-2025-00382.
}%
}

\begin{document}
\maketitle
\thispagestyle{empty}
\pagestyle{empty}

\begin{abstract}
An operator splitting method is developed for computing operating points of nonlinear resistive networks. Our behavioral formulation separates the device laws from Kirchhoff's laws and casts simulation as a zero-finding problem. Branch port resistances then provide a diagonal preconditioner for the relaxed Peaceman--Rachford iteration in scattering variables. Each iteration requires solving independent scalar equations and a linear system with a fixed weighted Laplacian matrix. Convergence is established for maximal monotone device laws when an operating point exists. Contractive branch reflections also yield a linear convergence bound independent of network topology. Numerical experiments on crossbar arrays illustrate the computational advantages.
\end{abstract}

\section{Introduction}

Analog computing has attracted renewed interest as a route to faster and more energy-efficient artificial intelligence~\cite{Momeni2025, ielmini2018memory,sebastian2020,huang2024memristor}.
Resistive crossbar arrays, for example, implement matrix-vector multiplication through Ohm's law and Kirchhoff's current law, computing where the data are stored and thereby reducing data movement~\cite{ielmini2018memory,sebastian2020,chen2019edge}. Related circuit architectures also support analog learning~\cite{scellier2024fast,chaffey2025circuit} and neuromorphic control~\cite{chaffey2024monotone,shahhosseini2025splitting,shahhosseini2025varmetric}. However, device variability, line resistances, and unwanted current paths limit the accuracy of analog computation~\cite{chen2013crossbar}. Accurate simulation of large-scale nonlinear resistive networks is therefore essential for circuit design.

General-purpose circuit simulators, such as SPICE~\cite{nagel1975spice2}, compute operating points by damped Newton iteration on the modified nodal analysis equations~\cite{ho1975mna,najm2010circuit}.  The device characteristics are linearized at the current iterate, and the resulting sparse Jacobian is factorized anew at each iteration. For large circuits, these repeated factorizations account for a substantial part of the computational cost~\cite{davis2010klu}, even though the topology is fixed and nonlinearities are local to each device.
This motivates an algorithm that reuses interconnection
computations while updating the nonlinear devices independently.

The paper develops a zero-finding formulation for computing operating points of nonlinear resistive networks and establishes its equivalence to a partial inverse problem~\cite{spingarn1983}. The formulation starts from the behavioral approach to systems theory~\cite{willems2007behavioral}, separating device laws from the constraints imposed by Kirchhoff's laws. The device laws define a possibly nonlinear, but separable product of branch behaviors, while Kirchhoff's laws define a linear subspace determined by the circuit topology. Their intersection is the circuit behavior, whose elements are the operating points. This separation allows a splitting method to exploit both the local nature of the nonlinearities and the fixed linear structure of the interconnection.

The main contribution is a circuit simulation algorithm obtained by leveraging the relaxed Peaceman--Rachford iteration~\cite{davis2016rates}. The iteration is expressed in scattering variables, with branch port resistances providing a diagonal preconditioner. Each iteration requires solving independent scalar equations and a linear system with a fixed weighted Laplacian matrix, whose factorization can be reused. Suitable choices of the relaxation parameter recover the Peaceman--Rachford and Douglas--Rachford iterations~\cite{lions1979,eckstein1992}. The latter also coincides with a preconditioned form of Spingarn's method of partial inverses~\cite{spingarn1983,mahey1995proximal}. This identification uses the orthogonality of admissible branch voltage and current spaces offered by Tellegen's theorem~\cite{tellegen1952,anderson1973network}. Convergence to an operating point, whenever one exists, is established for maximal monotone device laws using standard monotone operator theory~\cite{bauschke2017convex,davis2016rates,eckstein1992}. A linear convergence bound independent of network topology is also derived under the assumption that every branch reflected resolvent is contractive. 

\emph{Related work.} The problem of computing operating points of nonlinear resistive networks dates back to~\cite{minty1960}, and its formulation in terms of monotone relations is developed in~\cite{rockafellar1995monotone}. More recent operator approaches include maximal monotone one-ports~\cite{chaffey2024monotone}, a monotone plus skew formulation solved by primal-dual splitting, with a skew interconnection and step sizes constrained by the incidence matrix norm~\cite{chaffey2023monotoneskew}, operator classes beyond monotonicity~\cite{quan2025srg}, and fixed and variable metric splittings for neuromorphic circuits~\cite{shahhosseini2025splitting,shahhosseini2025varmetric}. A fast simulation method for nonlinear resistive networks in analog learning is proposed in~\cite{scellier2024fast}. Operator splitting over system behaviors for control is studied in~\cite{padoan2025split}  and for dynamic optimization in~\cite{padoan2026dynamic}. The contraction factor and optimal metric for the reflected resolvent of a strongly monotone and Lipschitz map are derived in~\cite{giselsson2017linear}. The variables and the interconnection step of Section~\ref{sec:main} coincide with the scattering variables and adaptor of wave digital filter theory~\cite{fettweis1986,bernardini2019}, whose fixed point iterations with several nonlinearities have been recently studied in~\cite{bernardini2025}.

\emph{Paper organization.} The paper is organized as follows. Section~\ref{sec:prelim} reviews monotone maps and the relaxed Peaceman-Rachford iteration. Section~\ref{sec:circuit} models resistive circuits in behavioral terms and defines the circuit simulation problem. Section~\ref{sec:main} derives the circuit simulation algorithm and analyzes its convergence. Section~\ref{sec:crossbar} presents a simulation case study of a crossbar array. Section~\ref{sec:conclusion} concludes the paper.

\section{Preliminaries}\label{sec:prelim}

We assume familiarity with monotone operator theory~\cite{bauschke2017convex}.
The notation largely follows~\cite{ryu2022}, with the standard
identification between set-valued maps and their graphs. Given  an inner product   $\langle \cdot,\cdot \rangle$  on $\mathbb{R}^n$,
with corresponding induced norm $\|\cdot\|$, a  (set-valued)  map ${T:\R^{n}\rightrightarrows\R^{n}}$ is \emph{monotone} if
\begin{equation*}
  \langle u-u',y-y'\rangle\ge0\quad\text{for all }(u,y),(u',y')\in\gra T,
\end{equation*}
and \emph{maximal monotone} if $\gra T$ is not properly contained in the graph of a monotone map \cite{bauschke2017convex,ryu2022}. 
A continuous, single-valued, monotone map ${T}$ on ${\R^n}$ is maximal monotone~\cite{bauschke2017convex}. 
The \emph{resolvent} and the \emph{reflected resolvent} of a  map $T$ are 
\begin{equation*}
  J_{T}=(I+T)^{-1},\ \text{ and } \ C_{T}=2J_{T}-I,
\end{equation*} 
\noindent
where $I$ is the identity map.
By Minty's theorem \cite{bauschke2017convex}, a monotone map $T$ is maximal monotone if and only if ${\dom J_{T}=\R^{n}}$, in which case $J_{T}$ is both single-valued and firmly nonexpansive, and $C_{T}$ is nonexpansive \cite{bauschke2017convex,ryu2022}. For a nonempty, closed, convex set ${\set{C}\subseteq\R^n}$,
the \emph{projection} onto $\set{C}$ is the map
${P_{\set{C}}:\R^n\to\set{C}}$ defined by ${u\mapsto\operatorname*{arg}\operatorname*{min}\limits_{y\in\set{C}}\|u-y\|}$, 
and the \emph{normal cone} of $\set{C}$ is the map
${N_{\set{C}}:\R^n\rightrightarrows\R^n}$ defined by
\begin{equation*}
  N_{\set{C}}(u)=\begin{cases}
  \{y\mid\langle y,z-u\rangle\le0\ \text{for all }z\in\set{C}\}, & u\in\set{C},\\
  \emptyset, & u\notin\set{C}.
  \end{cases}
\end{equation*}
The map $N_{\set{C}}$ is maximal monotone and ${J_{N_{\set{C}}}=P_{\set{C}}}$ \cite{bauschke2017convex}. If ${\V}$ is a subspace  of $\R^n$,  then ${N_{\V}(u)=\V^{\perp}}$ for every ${u\in\V}$, $P_{\V}$ is the orthogonal projection onto $\V$, and ${C_{N_{\V}}=2P_{\V}-I}$ is the reflection across $\V$. A product $T_1\times\cdots\times T_m$ of maps with nonempty graphs is maximal monotone if and only if $T_1,\dots,T_m$ are maximal monotone \cite{bauschke2017convex}.

Consider maximal monotone maps $F,G$ on $\R^{n}$ and the 
zero-finding problem of determining $u^{\star}\in\zer(F+G)$, 
that is, a solution of the inclusion
\begin{equation}
  0\in F(u)+G(u).
  \label{eq:sum}
\end{equation}
A widely used algorithm to solve \eqref{eq:sum} is the \emph{relaxed Peaceman-Rachford (PR) iteration} \cite{davis2016rates}
\begin{equation}
  a^{k+1}=(1-\alpha)\,a^{k}+\alpha\,C_{F}(C_{G}(a^{k})),\qquad k\in\Z_{+},
  \label{eq:DR}
\end{equation}
with ${\alpha\in(0,1]}$ the \emph{relaxation parameter}  and ${a^{0}\in\R^{n}}$ the initial condition. For ${\alpha=\tfrac12}$, \eqref{eq:DR} is the \emph{Douglas-Rachford (DR) iteration}, and for ${\alpha=1}$, it is the \emph{PR iteration} \cite{lions1979}. The zeros of $F+G$ and the fixed points of $C_{F}C_{G}$ are related by \cite{bauschke2017convex}
\begin{equation}
  \zer(F+G)=J_{G}(\fix C_{F}C_{G}).
  \label{eq:zerfix}
\end{equation}
The following convergence result is classical \cite{bauschke2017convex,davis2016rates,eckstein1992}.

\begin{theorem}[Convergence of the relaxed PR iteration]\label{thm:DR}
Consider maximal monotone maps $F,G$ on $\R^{n}$, ${\alpha\in(0,1)}$ and ${a^{0}\in\R^{n}}$. Assume ${\zer(F+G)\neq\emptyset}$. Then the sequence ${\{a^{k}\}_{k\in\Z_{+}}}$ generated by \eqref{eq:DR} converges to a point ${a^{\star}\in\fix C_{F}C_{G}}$, and ${\{J_{G}(a^{k})\}_{k\in\Z_{+}}}$ converges to ${J_{G}(a^{\star})\in\zer(F+G)}$.
\end{theorem}
\noindent
For ${\alpha=1}$, the iteration need not converge \cite{bauschke2017convex}. However, for every ${\alpha\in(0,1]}$, \eqref{eq:DR} converges linearly whenever $C_{F}$ is nonexpansive and $C_{G}$ is a contraction, by Banach's fixed point theorem \cite{bauschke2017convex}, since $(1-\alpha)I+\alpha C_{F}C_{G}$ is then a contraction.

\section{Problem formulation}\label{sec:circuit}

\subsection{Circuit topology and interconnection constraints}

The topology of a resistive network is naturally described by a connected, directed multigraph ${\bb{G}=(\Vertex_0,\E)}$, with node (or vertex) set ${\Vertex_0=\{0,1,\dots,n\}}$ and branch (or edge) set ${\E=\{1,\dots,m\}}$, in which parallel edges are allowed \cite{bollobas1998modern,bullo2022lectures}. Every two-terminal device of the circuit, including every source, is associated with a branch ${e\in\E}$ and the \emph{ground node} is taken to be ${0\in\Vertex_0}$, so that ${\Vertex=\{1,\dots,n\}}$ is the set of remaining nodes. An orientation of the branches is fixed arbitrarily \textit{a priori}. Each branch ${e\in\E}$ is associated with a voltage ${v_e\in\R}$ and a current ${i_e\in\R}$, with reference polarity and reference direction in accordance with the orientation of the branch~\cite{chua1987linear}. The vectors ${v,i\in\R^{m}}$ collect all branch voltages and the branch currents. Grounding the reference node and deleting the corresponding row
of the incidence matrix of $\bb{G}$ yields the
\emph{reduced incidence matrix} ${A\in\R^{n\times m}}$, with entries
${A_{ke}=1}$ if branch $e$ is oriented away from node $k$, ${A_{ke}=-1}$ if branch $e$ is oriented towards node $k$, and ${A_{ke}=0}$ otherwise, for all ${k\in\Vertex}$ and ${e\in\E}$. The matrix $A$ has full row rank because $\bb{G}$ is connected \cite[Ch.~2]{bollobas1998modern}, \cite{bullo2022lectures}, \cite[Ch.~4]{chua1987linear}. Denoting by ${\phi\in\R^{n}}$ the vector of potentials of the nodes in $\Vertex$, Kirchhoff's voltage law and Kirchhoff's current law read \cite[Ch.~4]{chua1987linear}
\begin{equation*}
  v=A^{\top}\phi,\qquad Ai=0.
\end{equation*}
Admissible voltages and currents thus define the subspaces
\begin{equation}
  \V=\ran A^{\top},\qquad \I=\ker A,
  \label{eq:kirchhoff}
\end{equation}
which, in the language of graph theory, are the \textit{cut space} and the \textit{cycle space} of $\bb{G}$ \cite[Ch.~2]{bollobas1998modern}. Note that ${\I=\V^{\perp}}$, since $\ker A$ is the orthogonal complement of ${\ran A^{\top}}$. Equivalently, ${v^{\top}i=0}$ for all ${v\in\V}$ and ${i\in\I}$, a fact known as Tellegen's theorem \cite{tellegen1952}, \cite[Ch.~2]{anderson1973network}. Since $A$ has full row rank, the node potentials are recovered from ${v\in\V}$ as ${\phi=(AA^{\top})^{-1}Av}$.

\subsection{Circuit behavior and constitutive relations}

 The circuit is modeled using the behavioral approach~\cite{willems2007behavioral},
with each device described by its set of admissible voltage-current pairs.  
The \emph{behavior} of branch ${e\in\E}$ is a nonempty, closed set ${\B_e\subseteq\R^{2}}$ of admissible voltage-current pairs $(v_e,i_e)$, and the \emph{branch relation} of branch $e$ is the set-valued map ${T_e:\R\rightrightarrows\R}$ with graph $\B_e$, that is,
\begin{equation}
  i_e\in T_e(v_e)\quad\text{if and only if}\quad(v_e,i_e)\in\B_e.
  \label{eq:branchrelation}
\end{equation}
By convention, current is taken as the dependent variable in~\eqref{eq:branchrelation}.
The devices considered in this paper are possibly nonlinear resistive elements, as well as constant voltage and current sources. Table~\ref{tab:behaviors} lists representative devices, the corresponding symbols used in circuit diagrams, their parameters, and their behaviors, each given as the set of admissible voltage-current pairs $(v_e,i_e)$ that satisfy the listed relation.

Behavioral systems theory models interconnection through variable sharing \cite{willems2007behavioral}. The \emph{device behavior} of the circuit is ${\B=\B_1\times\cdots\times\B_m}$, the set of pairs ${(v,i)\in\R^{m}\times\R^{m}}$ with ${(v_e,i_e)\in\B_e}$ for every ${e\in\E}$, and the \emph{branch relation of the circuit} is ${T=T_1\times\cdots\times T_m}$, so that ${i\in T(v)}$ if and only if ${(v,i)\in\B}$. The branches share their voltages and their currents through Kirchhoff's laws \eqref{eq:kirchhoff}, so that the \emph{circuit behavior} is
\begin{equation}
  \Cc=\B\cap(\V\times\I).
  \label{eq:circuit}
\end{equation}

\noindent
The elements of $\Cc$ are the \emph{operating points} of the circuit. The description \eqref{eq:circuit} of the operating points is a behavioral counterpart of the sparse tableau formulation of circuit analysis~\cite{hachtel1971sparse}, \cite[Ch.~5]{chua1987linear}, in which the device laws and Kirchhoff's laws are kept as separate sets of equations.

\subsection{Circuit simulation}

The \emph{circuit simulation problem} is to find an operating point ${(v,i)\in\Cc}$. Such a point need not exist or be unique. For example, two ideal voltage sources connected in parallel have no operating point if their voltages differ and infinitely many if they agree. Classical conditions for existence and uniqueness in monotone resistive networks are given in~\cite{minty1960}.

The circuit model~\eqref{eq:circuit} does not require monotone branch relations. Section~\ref{sec:main} develops an algorithm for computing an operating point. The resulting iteration is well defined for arbitrary initialization when each branch resolvent is single-valued and everywhere defined. Maximal monotonicity of the branch relations guarantees such properties and, with suitable parameters, convergence to an operating point when one exists.

\begin{table}[t]
\centering
\caption{Device symbols, parameters and behaviors.}
\label{tab:behaviors}
\small
\renewcommand{\arraystretch}{1.5}
\setlength{\tabcolsep}{4pt}
\begin{tabular}{@{}lcll@{}}
\toprule
Device & Symbol & Parameters & Behavior\\
\midrule
resistor &
\begin{circuitikz}[baseline=-0.1cm,scale=0.6,transform shape]
  \draw (0,0) to[R] (1.4,0);
\end{circuitikz}
& ${R>0}$ & ${v_e=Ri_e}$\\

voltage source &
\begin{circuitikz}[baseline=-0.1cm,scale=0.6,transform shape]
  \draw (1.4,0) to[american voltage source] (0,0);
\end{circuitikz}
& ${\bar{v}\in\R}$ & ${v_e=\bar{v}}$\\

current source &
\begin{circuitikz}[baseline=-0.1cm,scale=0.6,transform shape]
  \draw (0,0) to[american current source] (1.4,0);
\end{circuitikz}
& ${\bar{\imath}\in\R}$ & ${i_e=\bar{\imath}}$\\

ideal diode &
\begin{circuitikz}[baseline=-0.1cm,scale=0.6,transform shape]
  \draw (0,0) to[D] (1.4,0);
\end{circuitikz}
& & ${v_e\le0\le i_e,\ v_ei_e=0}$\\

Shockley diode &
\begin{circuitikz}[baseline=-0.1cm,scale=0.6,transform shape]
  \draw (0,0) to[D*] (1.4,0);
\end{circuitikz}
& ${I_s,V_T>0}$ &
${i_e=I_s(e^{v_e/V_T}-1)}$\\

nonlinear resistor &
\begin{circuitikz}[baseline=-0.1cm,scale=0.6,transform shape]
  \draw (0,0) to[generic] (1.4,0);
\end{circuitikz}
& ${g:\R\to\R}$ & ${i_e=g(v_e)}$\\
\bottomrule
\end{tabular}
\end{table}

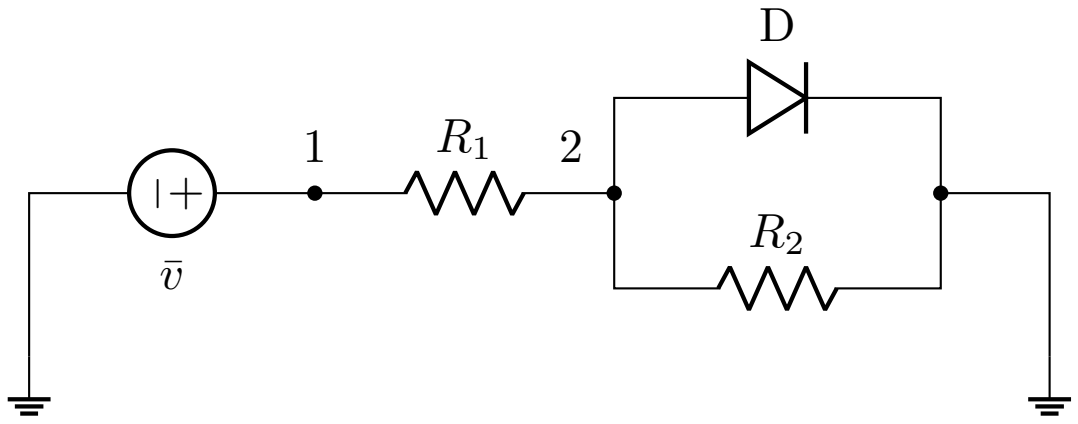
\begin{figure}[!t]
\centering
\begin{circuitikz}[scale=0.9,transform shape]
  \ctikzset{bipoles/length=1.05cm}

  \coordinate (n1) at (2.1,0);
  \coordinate (n2) at (4.3,0);
  \coordinate (g)  at (6.7,0);

  \draw (n1) to[american voltage source,l=$\bar{v}$]
        (0,0) -- (0,-1.2) node[ground]{};
  \draw (n1) to[R,l=$R_1$] (n2);

  \draw (n2) -- (4.3,0.7)
        to[D,l=$\mathrm{D}$] (6.7,0.7) -- (g);
  \draw (n2) -- (4.3,-0.7)
        to[R,l=$R_2$] (6.7,-0.7) -- (g);

  \draw (g) -- (7.5,0) -- (7.5,-1.2) node[ground]{};

  \fill (n1) circle (1.6pt);
  \fill (n2) circle (1.6pt);
  \fill (g) circle (1.6pt);
  \node[above=3pt] at (n1) {$1$};
  \node[above left=3pt] at (n2) {$2$};
\end{circuitikz}
\caption{Positive shunt diode clipper.}
\label{fig:example}
\end{figure}

\begin{example}[Positive shunt diode clipper]\label{ex:clipper}
Consider the positive shunt diode clipper illustrated in
Fig.~\ref{fig:example}, 
in which a voltage source $\bar{v}$ drives the load $R_2>0$ through
a current-limiting resistor $R_1>0$. An ideal diode
$\mathrm{D}$ connected in parallel with the load ``clips''
positive output voltages at zero.
The circuit has ${n=2}$ nodes besides the ground node and ${m=4}$ branches,  
and its reduced incidence matrix is
\begin{equation*}
  A=\begin{bmatrix}1&1&0&0\\0&-1&1&1\end{bmatrix},
\end{equation*}
so that the subspaces \eqref{eq:kirchhoff} are
\begin{subequations}\label{eq:exVI}
\begin{align}
  \V&=\{v\in\R^{4}\mid v_1=v_2+v_3,\ v_3=v_4\},\label{eq:exV}\\
  \I&=\{i\in\R^{4}\mid i_1+i_2=0,\ i_2=i_3+i_4\},\label{eq:exI}
\end{align}
\end{subequations}
which express Kirchhoff's voltage law around the two loops and Kirchhoff's current law at the two nodes. By Table~\ref{tab:behaviors}, the device behavior $\B$ is the set of pairs ${(v,i)\in\R^{4}\times\R^{4}}$ with
\begin{equation}
  v_1=\bar{v},\  v_2=R_1i_2,\  v_3\le0\le i_3,\, v_3i_3=0,\  v_4=R_2i_4,
  \label{eq:exB}
\end{equation}
and the circuit behavior \eqref{eq:circuit} is the set of pairs $(v,i)$ that satisfy \eqref{eq:exVI} and \eqref{eq:exB}. For $\bar{v}\le 0$, the diode does not conduct and the resistors form a voltage divider. For $\bar{v}>0$, the diode conducts, so that $v_3=v_4=0$ and  $i_3=\tfrac{\bar{v}}{R_1}$. For every $\bar{v}\in\R$, the circuit behavior is therefore a singleton. The output voltage across the load $R_2$ is
\begin{equation}
  v_3=v_4=\min\left\{\frac{R_2}{R_1+R_2}\,\bar{v},\,0\right\},
  \label{eq:exout}
\end{equation}
and ${v_2=\bar{v}-v_3}$, ${i_2=-i_1=\tfrac{v_2}{R_1}}$, ${i_4=\tfrac{v_3}{R_2}}$, ${i_3=i_2-i_4}$.
\end{example}

\section{Main results}\label{sec:main}

\subsection{Circuit simulation as a zero-finding problem}

Our modeling assumptions provide a natural separation between
the device laws and the interconnection constraints.
Circuit simulation thus amounts to finding a zero of the sum of
the branch relation $T$ and the normal cone $N_{\V}$, that is,
\begin{equation}
  0\in N_{\V}(v)+T(v).
  \label{eq:circuit-inclusion}
\end{equation}
The following result proves this equivalence and characterizes the operating points of the circuit in terms of solutions of~\eqref{eq:circuit-inclusion}.

\begin{proposition}[Operating points as zeros]\label{prop:zeros}
Consider the circuit behavior \eqref{eq:circuit} and let $T$ be its branch relation. Then 
\begin{equation}
  \zer(N_{\V}+T)=\dom\Cc,
  \label{eq:zero}
\end{equation}
where $\Cc$ is regarded as a set-valued map with the current as the dependent variable, as in \eqref{eq:branchrelation}. Moreover, ${(v,i)\in\Cc}$ if and only if ${v\in\zer(N_{\V}+T)}$ and ${i\in T(v)\cap\I}$.
\end{proposition}
\begin{proof}
By~\eqref{eq:kirchhoff}, ${\V^{\perp}=\I}$, so
$N_{\V}(v)=\I$ on $\V$ and $N_{\V}(v)=\emptyset$ elsewhere.
Since $\I$ is a subspace, \eqref{eq:circuit-inclusion}
holds if and only if ${v\in\V}$ and
${T(v)\cap\I\neq\emptyset}$.
By definition of $T$ and $\Cc$, this is equivalent to
${v\in\dom\Cc}$, proving~\eqref{eq:zero}.
For ${v\in\dom\Cc}$, ${(v,i)\in\Cc}$ if and only if
${i\in T(v)\cap\I}$, which proves the remaining claim.
\end{proof}

\noindent
Proposition~\ref{prop:zeros} shows that circuit simulation amounts
to finding ${v\in\V}$ and ${i\in\V^{\perp}}$ such that
${i\in T(v)}$, which is Spingarn's classical \emph{partial inverse
problem} for the pair $(T,\V)$~\cite{spingarn1983}. The
inclusion~\eqref{eq:circuit-inclusion} has the form \eqref{eq:sum},
with ${F=N_{\V}}$ and ${G=T}$ representing the interconnection
structure and the branch laws, respectively. Our circuit simulation algorithm is obtained by applying the relaxed PR iteration~\eqref{eq:DR} to a preconditioned inclusion equivalent to~\eqref{eq:circuit-inclusion}, as discussed next.

\subsection{Preconditioning by port resistances}

We introduce branch port resistances to precondition~\eqref{eq:circuit-inclusion} and tune the corresponding convergence rate.
Let ${\gamma_e>0}$ be the \emph{port resistance} assigned to
branch ${e\in\E}$, and let
${\Gamma=\diag(\gamma_1,\dots,\gamma_m)}$.
The associated inner product and induced norm on $\R^m$ are
\begin{equation}
  \langle x,y\rangle_{\Gamma}=x^{\top}\Gamma^{-1}y,\qquad
  \|x\|_{\Gamma}=\sqrt{x^{\top}\Gamma^{-1}x}.
  \label{eq:innerproduct}
\end{equation}
By Tellegen's theorem, $\Gamma\I$ is the orthogonal complement
of $\V$ in the inner product~\eqref{eq:innerproduct}.
Thus every ${a\in\R^m}$ decomposes uniquely as
${a=v+\Gamma i}$, with ${v\in\V}$ and ${i\in\I}$.
Moreover, since $\Gamma$ is invertible, \eqref{eq:circuit-inclusion}
is equivalent to the inclusion
\begin{equation}\label{eq:inclusion}
  0\in N_{\Gamma}(v)+T_{\Gamma}(v),
\end{equation}
where ${N_{\Gamma}=\Gamma N_{\V}}$ and ${T_{\Gamma}=\Gamma T}$. 

Our circuit simulation algorithm is obtained by applying the
relaxed PR iteration~\eqref{eq:DR} to the scaled
inclusion~\eqref{eq:inclusion}, giving
\begin{subequations}\label{eq:iter}
\begin{align}
  b^{k}&=C_{T_{\Gamma}}(a^{k}),\label{eq:iterb}\\
  a^{k+1}&=(1-\alpha)\,a^{k}+\alpha\,C_{N_{\Gamma}}(b^{k}),\label{eq:itera}
\end{align}
\end{subequations}
with ${k\in\Z_{+}}$ and ${a^{0}\in\R^{m}}$. 
The voltages and the currents are determined by $a^{k}$ and $b^{k}$ as
\begin{subequations}\label{eq:readout}
\begin{align}
  v^{k}&=\tfrac12(a^{k}+b^{k}),\label{eq:readoutv}\\
  i^{k}&=\tfrac12\Gamma^{-1}(a^{k}-b^{k}),\label{eq:readouti}
\end{align}
\end{subequations}
so that ${v^{k}=J_{T_{\Gamma}}(a^{k})}$, ${i^{k}=\Gamma^{-1}(a^{k}-v^{k})\in T(v^{k})}$ and ${(v^{k},i^{k})\in\B}$ for every ${k\in\Z_{+}}$. The iteration acts on $a^{k}$ and $b^{k}$ rather than on $(v^{k},i^{k})$, since $J_{T_{\Gamma}}$ is single-valued even when the branch relations are not (e.g., for sources and ideal diodes).

\begin{remark}[Scattering and passivity interpretation]\label{rem:scattering}
The variables ${a=v+\Gamma i}$ and ${b=v-\Gamma i}$ have
a classical interpretation in terms of scattering
\cite{belevitch1968classical,fettweis1986} and passivity theory
\cite{willems1972dissipative,vanderschaft2017l2gain}.
They represent the \emph{incident} and the \emph{reflected waves}
at the ports of the interconnection, with $\gamma_e$ the reference
resistance of port $e$, and the Lipschitz constant of the map
${a_e\mapsto b_e}$ is the \emph{reflection coefficient} of branch $e$.
The identity ${a_e^{2}-b_e^{2}=4\gamma_e\,v_ei_e}$ relates
the wave amplitudes to the power absorbed by branch ${e\in\E}$.
In the norm~\eqref{eq:innerproduct}, the corresponding identity is
$${\|a\|_{\Gamma}^{2}-\|b\|_{\Gamma}^{2}=4v^{\top}i}.$$
This identity is the scattering form of passivity
\cite{willems1972dissipative,vanderschaft2017l2gain}, so that
a branch absorbs power if and only if the amplitude of its
reflected wave is not larger than that of its incident wave.
\end{remark}

\subsection{Resolvents of the scaled maps}

Evaluating the scaled resolvents requires computing a weighted projection for the interconnection and solving independent scalar equations for the devices. Let
\begin{equation}
  P_{\V}=A^{\top}(A\Gamma^{-1}A^{\top})^{-1}A\Gamma^{-1}
  \label{eq:PGamma}
\end{equation}
denote the orthogonal projection onto ${\V=\ran A^{\top}}$ in the inner product~\eqref{eq:innerproduct}~\cite{golub2013matrix}. The following lemma characterizes the resolvents and maximal monotonicity properties of the scaled maps $N_{\Gamma}$ and $T_{\Gamma}$.

\begin{lemma}[Resolvents of $N_{\Gamma}$ and $T_{\Gamma}$]\label{lem:maps}
Consider the circuit behavior \eqref{eq:circuit} and port resistances ${\gamma_e>0}$,  with ${e\in\E}$, and equip $\R^{m}$ with the inner product and the induced norm \eqref{eq:innerproduct}. Then the following statements hold.
\begin{enumerate}[(i)]
\item ${N_{\Gamma}}$ coincides with the normal cone map of $\V$. Hence, $N_{\Gamma}$ is maximal monotone, ${J_{N_{\Gamma}}=P_{\V}}$ and ${C_{N_{\Gamma}}=2P_{\V}-I}$.
\item ${T_{\Gamma}}$ is maximal monotone if and only if $T$ is maximal monotone, and its resolvent is
\begin{equation}
  J_{T_{\Gamma}}(a)=\bigl(J_{\gamma_1T_1}(a_1),\dots,J_{\gamma_mT_m}(a_m)\bigr),
  \label{eq:branchwise}
\end{equation}
where ${J_{\gamma_eT_e}=(1+\gamma_eT_e)^{-1}}$ is the resolvent of $\gamma_eT_e$.
\end{enumerate}
\end{lemma}
\begin{proof}
(i) Since
${\langle z,v\rangle_{\Gamma}=(\Gamma^{-1}z)^{\top}v}$,
the orthogonal complement of $\V$ in~\eqref{eq:innerproduct}
is $\Gamma\I$. Thus $N_{\Gamma}$ is the normal cone of $\V$
in this inner product. The remaining claims follow from
Section~\ref{sec:prelim} and~\eqref{eq:PGamma}.

(ii) The identity
${\langle u-u',\Gamma(y-y')\rangle_{\Gamma}
=(u-u')^{\top}(y-y')}$
shows that the bijection ${(u,y)\mapsto(u,\Gamma y)}$
preserves monotonicity in the respective inner products
and graph inclusions, hence maximal monotonicity.
Finally,
${T_{\Gamma}=(\gamma_1T_1)\times\cdots\times(\gamma_mT_m)}$,
so~\eqref{eq:branchwise} follows from
the definition of the resolvent.
\end{proof}

\noindent
Lemma~\ref{lem:maps} establishes that the scaled maps
$N_{\Gamma}$ and $T_{\Gamma}$ satisfy
the maximal monotonicity assumptions of the relaxed PR
iteration whenever the branch relations are maximal monotone.
The resolvent formulas then suggest a circuit implementation
of the iteration, with independent scalar branch solves
for the devices and a weighted projection for the
interconnection.

\begin{remark}[Isometry of $C_{N_{\Gamma}}$] \label{rem:isometry}
The interconnection reflection $C_{N_{\Gamma}}$ has a useful property
for convergence analysis. The projection $P_{\V}$ is
self-adjoint and idempotent in the inner product
\eqref{eq:innerproduct}, so $C_{N_{\Gamma}}=2P_{\V}-I$
is self-adjoint with ${C_{N_{\Gamma}}^{2}=I}$. Hence
$${  \|C_{N_{\Gamma}}x\|_{\Gamma}^{2}
  =\langle x,C_{N_{\Gamma}}^{2}x\rangle_{\Gamma}
  =\|x\|_{\Gamma}^{2}}$$
for every ${x\in\R^{m}}$. Thus $C_{N_{\Gamma}}$ is an isometry
and, in particular, nonexpansive. Composition with $C_{N_{\Gamma}}$  preserves any nonexpansiveness or contraction
bound for $C_{T_{\Gamma}}$, allowing convergence analysis to use bounds derived from the
individual branch relations. For example, a maximal monotone map $T_{\Gamma}$ has a contractive
reflected resolvent $C_{T_{\Gamma}}$ if and only if both $T_{\Gamma}$ and
$T_{\Gamma}^{-1}$ are strongly monotone
\cite[Cor.~4.7]{bauschke2012firmly}, in which case~\eqref{eq:iter} converges globally and linearly for every
$\alpha\in(0,1]$.
\end{remark}

\subsection{Circuit simulation algorithm}

The resolvent formulas yield an efficient implementation
of~\eqref{eq:iter}. By Lemma~\ref{lem:maps}, the auxiliary variable $b^{k}$ may be computed one branch at a time, as ${b^{k}_e=2J_{\gamma_eT_e}(a^{k}_e)-a^{k}_e}$. By \eqref{eq:PGamma}, \eqref{eq:itera} requires ${P_{\V}(b^{k})=A^{\top}\phi^{k}}$, where ${\phi^{k}\in\R^{n}}$ solves the linear system
\begin{equation}
  A\Gamma^{-1}A^{\top} \phi^{k}=A\Gamma^{-1}b^{k}.
  \label{eq:nodal}
\end{equation}
The resulting procedure is summarized in Algorithm~\ref{alg:bis}.

\noindent
\begin{minipage}{\columnwidth}
\refstepcounter{algo}\label{alg:bis}
\hrule\vspace{2pt}
\textbf{Algorithm~\thealgo}\ \ Behavioral circuit simulation\vspace{2pt}\hrule\vspace{3pt}
\noindent
\textbf{Input:} reduced incidence matrix ${A\in\R^{n\times m}}$, branch relations ${T_e:\R\rightrightarrows\R}$ and port resistances ${\gamma_e>0}$ for ${e\in\E}$, relaxation parameter ${\alpha\in(0,1]}$, initial condition ${a^{0}\in\R^{m}}$.
\begin{enumerate}
\item[1:] \textbf{for} ${k=0,1,\ldots}$ \textbf{do}
\item[2:] \quad ${b^{k}_e=2J_{\gamma_eT_e}(a^{k}_e)-a^{k}_e}$ for every ${e\in\E}$
\item[3:] \quad solve ${A\Gamma^{-1}A^{\top}\phi^{k}=A\Gamma^{-1}b^{k}}$ for ${\phi^{k}\in\R^{n}}$
\item[4:] \quad ${a^{k+1}=(1-\alpha)\,a^{k}+\alpha\,(2A^{\top}\phi^{k}-b^{k})}$
\item[5:] \textbf{end for}
\end{enumerate}
\textbf{Output:} sequences ${\{a^{k}\}_{k\in\Z_{+}}}$ and ${\{b^{k}\}_{k\in\Z_{+}}}$. 
\vspace{2pt}\hrule
\end{minipage}
\medskip

\noindent
Algorithm~\ref{alg:bis} is the relaxed PR iteration \eqref{eq:DR} for the pair $(N_{\Gamma},T_{\Gamma})$ in the inner product \eqref{eq:innerproduct}.  For ${\alpha=\tfrac12}$ and ${\gamma_e=1}$ for every ${e\in\E}$, it reduces to Spingarn's method of partial inverses for the pair $(T,\V)$ \cite{spingarn1983}, an identification due to \cite[\S5]{eckstein1992}, and for general port resistances it is a scaled version of the method studied in~\cite{mahey1995proximal}. By \eqref{eq:zerfix} and Proposition~\ref{prop:zeros}, $J_{T_{\Gamma}}$ maps fixed points of $C_{N_{\Gamma}}C_{T_{\Gamma}}$ onto voltages of operating points. The voltages and currents $v^{k}$ and $i^{k}$ are recovered from $a^{k}$ and $b^{k}$
using~\eqref{eq:readout}.

\begin{remark}[Computational aspects]\label{rem:cost}
The matrix $L=A\Gamma^{-1}A^{\top}$ in~\eqref{eq:nodal}
is the Laplacian matrix of $\bb{G}$ weighted by the port
conductances ${\{\gamma_e^{-1}\}_{e\in\E}}$~\cite{bullo2022lectures}, 
with the row and column of the
ground node removed. It is symmetric and positive definite,
since $\bb{G}$ is connected and $A$ has full row rank.
A sparse Cholesky factorization ${L=KK^{\top}}$ with a
fill-reducing ordering can be computed once, before the
iterations~\cite[Ch.~11]{golub2013matrix}, so each network
solve reduces to two sparse triangular solves. For large graphs, approximate Cholesky factorizations yield
solvers for~\eqref{eq:nodal} with nearly linear expected
running time in $m$ for a fixed relative
accuracy~\cite{kyng2016approximate}. 
The factorization depends only on the topology and the port resistances, so it is reused across iterations and across simulations of the same network with different source values. Since every iterate satisfies the branch relations, the iteration is stopped when the residuals ${\|Ai^{k}\|}$ and ${\|v^{k}-P_{\V}v^{k}\|}$ of Kirchhoff's laws fall below a given tolerance. Finally, the evaluation of~\eqref{eq:iterb} through~\eqref{eq:branchwise}, the update~\eqref{eq:itera} once $\phi^{k}$ is known, and the readouts~\eqref{eq:readout} are \emph{embarrassingly parallel} over the branches~\cite[Ch.~4]{ryu2022}.
\end{remark}

\subsection{Convergence analysis}\label{sec:conv}

The convergence of Algorithm~\ref{alg:bis} follows from Theorem~\ref{thm:DR} applied to the pair $(N_{\Gamma},T_{\Gamma})$, with rates determined by the branches alone. For every branch ${e\in\E}$ with maximal monotone branch relation $T_e$, let
${\rho_e \ge 0}$ be the Lipschitz constant of $C_{\gamma_eT_e}$, which is at most one by Section~\ref{sec:prelim} and is referred to as the \emph{reflection coefficient} of branch $e$, and let ${\bar{\rho}=\max_{e\in\E}\rho_e}$.

\begin{theorem}[Convergence of Algorithm~\ref{alg:bis}]\label{thm:main}
Consider the circuit behavior \eqref{eq:circuit}, port resistances ${\gamma_e>0}$ for ${e\in\E}$, and the sequences generated by Algorithm~\ref{alg:bis} from ${a^{0}\in\R^{m}}$. Assume that $T_e$ is maximal monotone for every ${e\in\E}$. Then all sequences are well defined, and the following statements hold.
\begin{enumerate}[(i)]
\item If ${\Cc\neq\emptyset}$ and ${\alpha\in(0,1)}$, then ${\{a^{k}\}_{k\in\Z_{+}}}$ converges to a point ${a^{\star}\in\fix C_{N_{\Gamma}}C_{T_{\Gamma}}}$, and ${\{(v^{k},i^{k})\}_{k\in\Z_{+}}}$ converges to an operating point ${(v^{\star},i^{\star})\in\Cc}$, with ${a^{\star}=v^{\star}+\Gamma i^{\star}}$.
\item If ${\bar{\rho}<1}$, then $\Cc$ is a singleton $\{(v^{\star},i^{\star})\}$, the point ${a^{\star}=v^{\star}+\Gamma i^{\star}}$ is the unique fixed point of $C_{N_{\Gamma}}C_{T_{\Gamma}}$, and for every ${\alpha\in(0,1]}$ and ${k\in\Z_{+}}$,
\begin{subequations}
\begin{align}
  \|a^{k+1}-a^{\star}\|_{\Gamma}&\le(1-\alpha+\alpha\bar{\rho})\,\|a^{k}-a^{\star}\|_{\Gamma},
  \label{eq:rate} \\
  \|v^{k}-v^{\star}\|_{\Gamma}&\le\|a^{k}-a^{\star}\|_{\Gamma}, \label{eq:readoutrate1}\\
  \|\Gamma(i^{k}-i^{\star})\|_{\Gamma}&\le\|a^{k}-a^{\star}\|_{\Gamma}.
  \label{eq:readoutrate2}
\end{align}
\end{subequations}
\end{enumerate}
\end{theorem}

\begin{proof}
By Lemma~\ref{lem:maps} and Minty's theorem, all sequences
are well defined. Define the map
${S:\R^{m}\to\R^{m}\times\R^{m}}$ by
\[
  S(a)=\bigl(J_{T_{\Gamma}}(a),
    \Gamma^{-1}(a-J_{T_{\Gamma}}(a))\bigr).
\]
For ${(v,i)=S(a)}$, we have ${i\in T(v)}$ and
\begin{align*}
  C_{N_{\Gamma}}C_{T_{\Gamma}}(a)=a
  &\iff P_{\V}(v-\Gamma i)=v\\
  &\iff v\in\V,\quad i\in\I.
\end{align*}
Thus $S$ restricts to a bijection from
${\fix C_{N_{\Gamma}}C_{T_{\Gamma}}}$ onto $\Cc$,
with inverse ${(v,i)\mapsto v+\Gamma i}$.

(i) Since
${\zer(N_{\Gamma}+T_{\Gamma})=\zer(N_{\V}+T)=\dom\Cc\neq\emptyset}$
by Proposition~\ref{prop:zeros}, Theorem~\ref{thm:DR} gives
${a^{k}\to a^{\star}\in\fix C_{N_{\Gamma}}C_{T_{\Gamma}}}$.
By continuity of $S$,
${(v^{k},i^{k})=S(a^{k})\to S(a^{\star})\in\Cc}$.

(ii) By~\eqref{eq:branchwise} and the diagonal form of $\Gamma$,
$C_{T_{\Gamma}}$ is $\bar{\rho}$-Lipschitz in
$\|\cdot\|_{\Gamma}$.
Since $C_{N_{\Gamma}}$ is an isometry
by Remark~\ref{rem:isometry},
$C_{N_{\Gamma}}C_{T_{\Gamma}}$ has the same bound,
and ${(1-\alpha)I+\alpha C_{N_{\Gamma}}C_{T_{\Gamma}}}$
is a contraction with constant
${1-\alpha+\alpha\bar{\rho}<1}$.
Banach's fixed point theorem gives a unique fixed point
$a^{\star}$ and~\eqref{eq:rate}.
The bijection defined by $S$ then gives a unique operating point.
Finally, $J_{T_{\Gamma}}$ and ${I-J_{T_{\Gamma}}=J_{T_{\Gamma}^{-1}}}$ are nonexpansive \cite{bauschke2017convex}, which yields
\eqref{eq:readoutrate1} and~\eqref{eq:readoutrate2}.%
\end{proof}

\noindent
Theorem~\ref{thm:main} establishes convergence to an operating
point whenever one exists and guarantees existence, uniqueness
and linear convergence when every branch reflection is contractive.
Part (i) covers every circuit with maximal monotone branches that has an operating point, including circuits with ideal sources and ideal diodes, and the residual ${\|a^{k+1}-a^{k}\|_{\Gamma}}$ then vanishes as ${o(1/\sqrt{k})}$ \cite{davis2016rates}. The restriction ${\alpha<1}$ in (i) cannot be removed. A voltage source in parallel with a current source illustrates
this limitation. The map $C_{N_{\Gamma}}C_{T_{\Gamma}}$ acts
as a nontrivial rotation about its unique fixed point
$a^{\star}$ in the inner product \eqref{eq:innerproduct}.
Algorithm~\ref{alg:bis} with ${\alpha=1}$ therefore preserves
the distance to $a^{\star}$ and fails to converge from any
${a^{0}\neq a^{\star}}$. Part (ii) requires every reflection coefficient to be smaller than one. It then gives existence and uniqueness of the operating point, and a rate that does not depend on the graph. The bound \eqref{eq:rate} is smallest for ${\alpha=1}$, where it equals $\bar{\rho}$, so the PR iteration is the choice of reference when ${\bar{\rho}<1}$, whereas ${\alpha<1}$ is required when ${\bar{\rho}=1}$.  A theoretical characterization of how port resistances affect convergence if $\bar{\rho}=1$ and $\alpha<1$ is left for future work.

\subsection{Computing reflection coefficients and branch resolvents}\label{ssec:computing-branch-resolvents}

Algorithm~\ref{alg:bis} requires solving, for every branch
${e\in\E}$, the scalar equation ${a_e=v_e+\gamma_ei_e}$,
with ${(v_e,i_e)\in\B_e}$, whose solution
${v_e=J_{\gamma_eT_e}(a_e)}$ is unique by Minty's theorem
if $T_e$ is maximal monotone. Table~\ref{tab:resolvents}
reports the resolvents and the reflection coefficients of the
devices of Table~\ref{tab:behaviors}. Each resolvent may be
derived by checking that ${(v_e,i_e)\in\B_e}$, with
${i_e=\tfrac{a_e-v_e}{\gamma_e}}$, and each reflection
coefficient follows from the resolvent by direct computation.
For example, a resistor $R>0$ has resolvent
${J_{\gamma_eT_e}(a_e)=Ra_e/(R+\gamma_e)}$
and reflection coefficient $|R-\gamma_e|/(R+\gamma_e)$. A resistor $R$ in series with a voltage source $\bar{v}$, with behavior ${v_e=Ri_e+\bar{v}}$, has resolvent ${J_{\gamma_eT_e}(a_e)=(Ra_e+\gamma_e\bar{v})/(R+\gamma_e)}$ and the same reflection coefficient, since $C_{\gamma_eT_e}$ is affine with slope ${(R-\gamma_e)/(R+\gamma_e)}$.
Choosing ${\gamma_e=R}$ makes this coefficient zero and
corresponds to \emph{impedance matching}~\cite{fettweis1986}.

The ideal diode admits the explicit, nonexpansive resolvent
${J_{\gamma_eT_e}(a_e)=\min\{a_e,0\}}$, so its set-valued branch relation can be used
directly in the iteration. By contrast, Newton's method on
the nodal equations requires a differentiable characteristic,
and circuit simulators resort to damping and
continuation~\cite{najm2010circuit}.

For a nonlinear resistor, with ${g:\R\to\R}$ continuously
differentiable and increasing, let
\begin{equation}
  h_e:\R\to\R,\qquad h_e(v_e)=v_e+\gamma_eg(v_e).
  \label{eq:newtonfun}
\end{equation}
Since ${h_e'(v_e)=1+\gamma_eg'(v_e)\ge1}$ for every
${v_e\in\R}$, the function $h_e$ is strictly increasing
and onto $\R$, so ${J_{\gamma_eT_e}=h_e^{-1}}$, and
${v_e=h_e^{-1}(a_e)}$ may be computed by Newton's
method~\cite{najm2010circuit}, vectorized over all such
branches. For example, a diode may be modeled by the Shockley equation 
\begin{equation}
  i_e=g(v_e)
  =I_s\left(\exp\left(\frac{v_e}{V_T}\right)-1\right),
  \label{eq:shockley-diode}
\end{equation}
with ${I_s,V_T>0}$. Its resolvent is
${J_{\gamma_eT_e}(a_e)=h_e^{-1}(a_e)}$, with $h_e$ defined
by~\eqref{eq:newtonfun} for this choice of $g$.
An explicit resolvent calculation is given
in~\cite{chaffey2025circuit}.

For a nonlinear resistor satisfying
${g'\in[R_{\max}^{-1},R_{\min}^{-1}]}$
for some ${0<R_{\min}\le R_{\max}}$, the derivative
$${C'_{\gamma_eT_e}(a_e)=(1-\gamma_es)/(1+\gamma_es)}$$
decreases with ${s=g'(v_e)}$. Hence,
\begin{equation}
  \rho_e\le\max\Bigl\{
    \frac{|R_{\max}-\gamma_e|}{R_{\max}+\gamma_e},
    \frac{|R_{\min}-\gamma_e|}{R_{\min}+\gamma_e}
  \Bigr\}.
  \label{eq:rhosector}
\end{equation}
The right-hand side is minimized by
${\gamma_e=\sqrt{R_{\min}R_{\max}}}$, with minimum value
${(\sqrt{\kappa}-1)/(\sqrt{\kappa}+1)}$,
where ${\kappa=R_{\max}/R_{\min}}$.

\begin{table}[t!]
\centering
\caption{Device resolvents and reflection coefficients.}
\label{tab:resolvents}
\renewcommand{\arraystretch}{1.7}
\setlength{\tabcolsep}{4pt}
\begin{tabular}{@{}lcc@{}}
\toprule
Device & $J_{\gamma_eT_e}(a_e)$ & $\rho_e$\\
\midrule
resistor & ${\dfrac{R\,a_e}{R+\gamma_e}}$ & ${\dfrac{|R-\gamma_e|}{R+\gamma_e}}$\\
voltage source & $\bar{v}$ & $1$\\
current source & ${a_e-\gamma_e\bar{\imath}}$ & $1$\\
ideal diode & ${\min\{a_e,0\}}$ & $1$\\
Shockley diode & ${h_e^{-1}(a_e)}$, $g$ in~\eqref{eq:shockley-diode} & $1$\\
nonlinear resistor & ${h_e^{-1}(a_e)}$, $h_e$ in \eqref{eq:newtonfun} & \eqref{eq:rhosector}\\
\bottomrule
\end{tabular}
\end{table}

\section{Case Study: A Resistive Crossbar Array}\label{sec:crossbar}

Memristive crossbar arrays have attracted considerable attention as the substrate of analog in-memory computing and of neuromorphic hardware, since they store a matrix in the conductances of their devices and compute a matrix-vector product by applying voltages to their word lines and reading the currents of their bit lines \cite{ielmini2018memory,sebastian2020}. However, line resistances and sneak paths through unselected devices make the result differ from the ideal product, so that assessing an array requires its operating point at every input \cite{chen2013crossbar}. This section illustrates the performance of Algorithm~\ref{alg:bis} on the problem of simulating a resistive crossbar array in comparison with \texttt{ngspice}~\cite{ngspice}, an
open-source SPICE-based circuit simulator.

\subsection{Circuit modeling and port resistances}

\begin{figure}[t!]
\centering
\ctikzset{bipoles/length=0.8cm}
\begin{circuitikz}[scale=0.7,transform shape]
  \foreach \y/\lab in {4.0/1, 2.3/2, -0.6/N}{
    \draw (-1.9,\y) -- (-1.9,\y-0.45) node[ground]{};
    \draw (-1.9,\y)
  to[american voltage source,l_=$\bar{V}_{\lab}$,invert] (-0.5,\y)
  to[R] (0.9,\y) coordinate (w\lab1);
    \draw (w\lab1) -- (2.0,\y)
      to[R] (3.3,\y) coordinate (w\lab2);
    \draw (w\lab2) -- (4.0,\y);
    \node at (4.5,\y) {$\cdots$};
    \draw (5.0,\y) -- (5.7,\y) coordinate (w\lab N);
  }

  \node[font=\scriptsize,above] at (0.2,4.1) {$R_s$};
  \node[font=\scriptsize,above] at (2.65,4.1) {$R_\ell$};

  \foreach \x/\c in {0.9/1, 3.3/2, 5.7/N}{
    \foreach \y/\r in {4.0/1, 2.3/2, -0.6/p}{
      \draw (\x,\y)
        to[generic] (\x+0.9,\y-0.85) coordinate (b\r\c);
    }
    \draw (b1\c) to[R] (\x+0.9,2.4) -- (b2\c);
    \draw (b2\c) -- (\x+0.9,0.75);
    \node at (\x+0.9,0.25) {$\vdots$};
    \draw (\x+0.9,-0.25) -- (bp\c);
    \draw (bp\c) to[R] (\x+0.9,-2.9) node[ground]{};
  }

  \node[font=\scriptsize,right] at (6.65,2.8) {$R_\ell$};
  \node[font=\scriptsize,right] at (6.65,-2.2) {$R_L$};
  \node at (-1.2,0.9) {$\vdots$};

  \foreach \x in {0.9,3.3,5.7}{
    \foreach \y in {4.0,2.3,-0.6}{
      \fill (\x,\y) circle (1.6pt);
    }
  }
  \foreach \x in {1.8,4.2,6.6}{
    \foreach \y in {3.15,1.45,-1.45}{
      \fill (\x,\y) circle (1.6pt);
    }
  }

  \node[font=\scriptsize,above] at (0.9,4.1) {$w_{1,1}$};
  \node[font=\scriptsize,above] at (5.7,4.1) {$w_{1,N}$};
  \node[font=\scriptsize,right] at (6.65,-1.45) {$\beta_{N,N}$};
  \node[font=\scriptsize,right] at (1.85,3.15) {$\beta_{1,1}$};
\end{circuitikz}
\caption{Square crossbar array with $N$ word and $N$ bit lines. Each cross-point of the array is joined by either a nonlinear resistor alone (1R), or a resistor in series with a diode modeled either ideally (ideal 1D1R) or with the Shockley equation (Shockley 1D1R).}
\label{fig:xbar}
\end{figure}
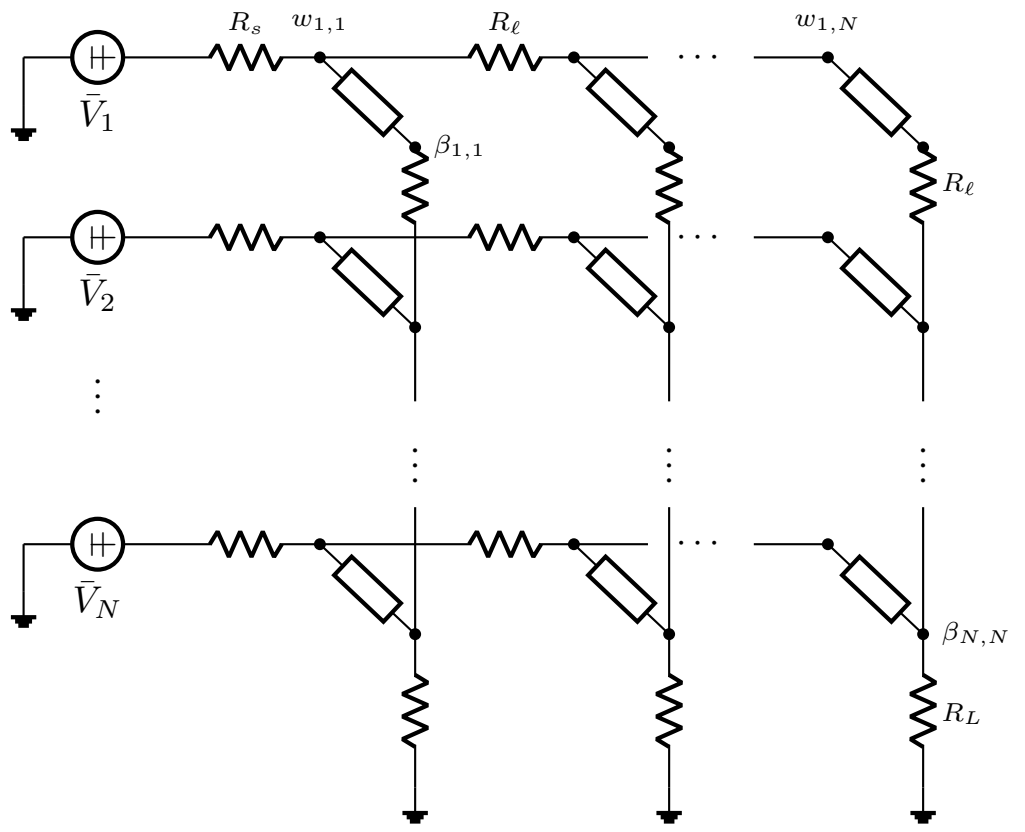

Consider the $N \times N$ crossbar array of Fig.~\ref{fig:xbar}, with $N$ word and $N$ bit lines. Word line $r$ is associated with the nodes $w_{r,1},\dots,w_{r,N}$, joined by line resistances $R_{\ell}$, and bit line $c$ is associated with the nodes $\beta_{1,c},\dots,\beta_{N,c}$, joined by the same line resistances. Each word line is driven at its head by a voltage source $\bar{V}_r$ with series resistance $R_{s}$, and each bit line is terminated at its foot by a load $R_{L}$. The source and its series resistance form one branch, whose resolvent and reflection coefficient are given in Table~\ref{tab:resolvents}. 
Several cross-point configurations have been studied for
resistive crossbar arrays, including one-resistor (1R) and
one-diode--one-resistor (1D1R)
cells~\cite{chen2013crossbar,Liu2019}.
We consider three circuit models for joining $w_{r,c}$ to $\beta_{r,c}$ based on these configurations.

Motivated by the piecewise-linear resistor characteristics
considered in~\cite{chua1987linear}, we first consider a
\emph{1R} model, where each cross-point is joined by a nonlinear resistor with branch relation
\begin{equation} \label{eq:nonlinear-resistor}
    i = \frac{v}{R_{\max}} + \left(\frac{1}{R_{\min}}  - \frac{1}{R_{\max}} \right) \tanh(v) .
\end{equation}
Next, we consider the \emph{ideal 1D1R} model~\cite{chen2013crossbar,Liu2019}, which places an ideal diode (with behavior in Table \ref{tab:behaviors}) in series with the 1R device. Finally, we consider the \emph{Shockley 1D1R} model, which places a diode governed by the Shockley equation~\eqref{eq:shockley-diode} in series with the 1R device.
For 1D1R arrays, the circuit has $n=3N^2$ nodes and $m=4N^2$ branches, namely $N(N-1)$ word line resistors, $N(N-1)$ bit line resistors, $2N^2$ device branches, $N$ source branches and $N$ loads. Similarly, 1R arrays have $n=2N^2$ nodes and $m=3N^2$ branches.

The line, source, and load branches are matched by setting
their port resistances equal to the corresponding
resistances. Each nonlinear resistor with branch relation \eqref{eq:nonlinear-resistor} has incremental resistance in ${[R_{\min},R_{\max}]}$, so that the port resistance minimizing the bound \eqref{eq:rhosector} is ${\gamma_e=\sqrt{R_{\min}R_{\max}}}$. However, voltage drops along the lines bias the operating point towards ${v_e=0}$, where the incremental resistance is close to $R_{\min}$. This motivates the choice ${\gamma_e=R_{\min}}$, which is used in our experiments. All diode models in Table~\ref{tab:resolvents} have reflection coefficient one. For diodes of both types, we use ${\gamma_e=\sqrt{R_{\min}R_{\max}}}$, which yields faster convergence in our  experiments.

\subsection{Numerical implementation and results}

We compare Algorithm~\ref{alg:bis} with \texttt{ngspice}
on the three crossbar arrays models described above of size $N\times N$, with $N = 16, 32, \dots, 1024$. For each array size, ten random circuits are generated and simulated with both methods. For each crossbar array, the line resistances are ${R_{\ell}=5\,\Omega}$, the source voltages $\bar{V}_r$ are drawn uniformly from ${(-1,1.5)}$\,V, the source resistances are ${R_s=50\,\Omega}$ and the loads are ${R_L=1\,\mathrm{k\Omega}}$. Every nonlinear resistor is described by \eqref{eq:nonlinear-resistor}, with $R_{\min}$ and $R_{\max}$
drawn per cell, log-uniformly, from $[1, 10]\,\mathrm{k\Omega}$ and $[50, 200]\,\mathrm{k\Omega}$. In the 1D1R arrays,
each resistor is in series with a diode, either using the ideal or Shockley model, with $I_s = 1$\,nA, $V_T = 25.85$\,mV. The resolvents of the nonlinear resistors~\eqref{eq:nonlinear-resistor}
and Shockley diodes~\eqref{eq:shockley-diode} are computed
numerically as described in
Section~\ref{ssec:computing-branch-resolvents}. 

Algorithm~\ref{alg:bis} starts from ${a^{0}=0}$, with relaxation parameter $\alpha = 1$ for 1R arrays and $\alpha = 0.8$ for 1D1R arrays.
It factorizes the weighted Laplacian $A\Gamma^{-1}A^{\top}$ once with CHOLMOD (AMD ordering, one thread). It stops when KCL holds at every node to relative tolerance $10^{-9}$ and absolute tolerance $10^{-12}$\,A, and the KVL residual is below $10^{-9}\max|v|$. Across all experiments the two solvers' node potentials agree to within $2.2\times10^{-7}$\,V on the 1R and ideal 1D1R arrays and $8\times10^{-6}$\,V on the
Shockley 1D1R arrays. All experiments were conducted on the same desktop computer (6-core Intel Core i5, 31.8\,GB RAM), each in a separate process. Runtime is measured as wall-clock time, including setup. Memory usage is measured as the peak process working set.

\begin{figure}[t]
  \centering
  \begin{tikzpicture}
\begin{groupplot}[group style={nm grid style=3, vertical sep=0.15cm, horizontal sep=0.12cm}, nm axis,
  width={(\columnwidth-1.3cm)/3}, height=1.9cm, xlabel={}, ylabel={},
  tick label style={font=\tiny}, title style={font=\scriptsize, yshift=1pt},
  xlabel style={font=\scriptsize, yshift=2pt},
  every axis plot/.append style={mark size=1.3pt, line width=0.7pt},
  /pgfplots/nm runtime row/.append style={ytick={1e-3,1e-1,1e1,1e3}},
  /pgfplots/nm iterations row/.style={ymin=8, ymax=3e3, ytick={1e1,1e2,1e3}},
  /pgfplots/nm memory row/.append style={ymin=0.05, ymax=40, ytick={0.1,1,10}}]
\nextgroupplot[nm runtime row, title={1R}, legend to name=nmRuntimeMemoryLegend, nm legend]
  \addplot[nm spice, dashed]     table[nm csv, y=mem_spice] {figures/network_models/runtime.csv}; \addlegendentry{SPICE}
  \addplot[nm zero bias] table[nm csv, y=mem_zb]       {figures/network_models/runtime.csv}; \addlegendentry{Algorithm \ref{alg:bis}}
\nextgroupplot[nm runtime row, title={Ideal 1D1R}]
  \addplot[nm spice, dashed]     table[nm csv, y=ideal_spice] {figures/network_models/runtime.csv};
  \addplot[nm zero bias] table[nm csv, y=ideal_zb]       {figures/network_models/runtime.csv};
\nextgroupplot[nm runtime row, title={Shockley 1D1R}]
  \addplot[nm spice, dashed]     table[nm csv, y=shockley_spice] {figures/network_models/runtime.csv};
  \addplot[nm zero bias] table[nm csv, y=shockley_zb]       {figures/network_models/runtime.csv};
\nextgroupplot[nm memory row]
  \addplot[nm spice, dashed]     table[nm csv, y=mem_spice] {figures/network_models/memory.csv};
  \addplot[nm zero bias] table[nm csv, y=mem_zb]       {figures/network_models/memory.csv};
\nextgroupplot[nm memory row]
  \addplot[nm spice, dashed]     table[nm csv, y=ideal_spice] {figures/network_models/memory.csv};
  \addplot[nm zero bias] table[nm csv, y=ideal_zb]       {figures/network_models/memory.csv};
\nextgroupplot[nm memory row]
  \addplot[nm spice, dashed]     table[nm csv, y=shockley_spice] {figures/network_models/memory.csv};
  \addplot[nm zero bias] table[nm csv, y=shockley_zb]       {figures/network_models/memory.csv};
\nextgroupplot[nm iterations row, xlabel={$N$}]
  \addplot[nm zero bias] table[nm csv, y=mem_zb] {figures/network_models/iterations.csv};
\nextgroupplot[nm iterations row, xlabel={$N$}]
  \addplot[nm zero bias] table[nm csv, y=ideal_zb] {figures/network_models/iterations.csv};
\nextgroupplot[nm iterations row, xlabel={$N$}]
  \addplot[nm zero bias] table[nm csv, y=shockley_zb] {figures/network_models/iterations.csv};
\end{groupplot}
\end{tikzpicture}
  \caption{Runtime in seconds (top), peak memory in GB (middle) and, for Algorithm \ref{alg:bis} only, iterations to convergence (bottom) for \texttt{ngspice} (dashed) and Algorithm \ref{alg:bis} (solid), median over 10 random circuits per size. On the 1R and ideal 1D1R arrays, the KLU factorization of \texttt{ngspice} fails at ${N=1024}$, and on the Shockley 1D1R arrays \texttt{ngspice} times out after $1600$\,s at ${N=512}$.}
  \label{fig:nm-runtime-memory}
\end{figure}
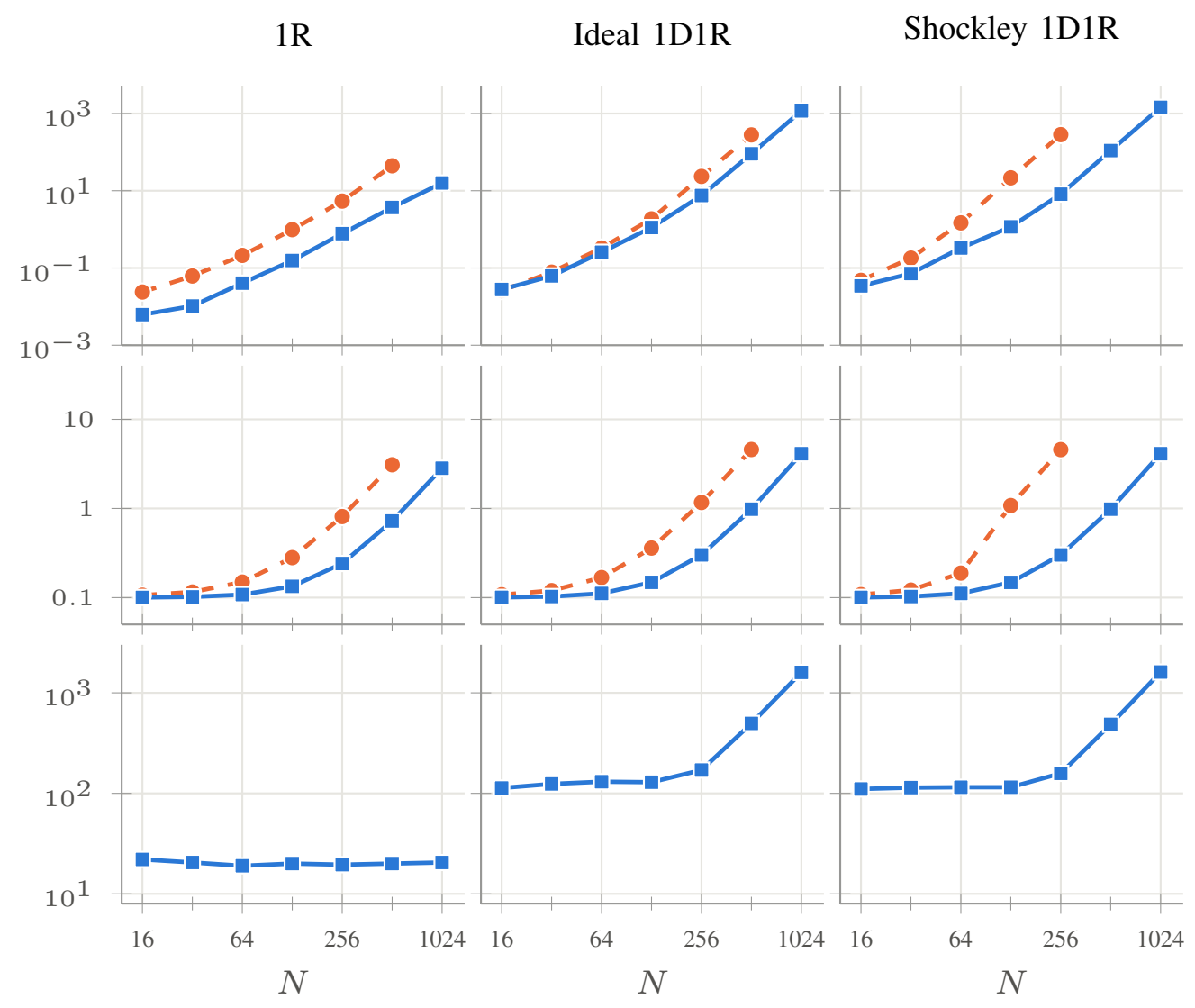

Fig.~\ref{fig:nm-runtime-memory} reports our numerical results. Algorithm \ref{alg:bis} is faster and requires less memory than \texttt{ngspice} in every run with $N>16$. On the 1R array, Algorithm \ref{alg:bis} converges in a number of iterations that does not depend on the array size, consistent with the bound of Theorem~\ref{thm:main}, which does not depend on the topology.
For the 1R and ideal 1D1R arrays, \texttt{ngspice}'s KLU factorization failed at $N=1024$. For the Shockley 1D1R arrays, \texttt{ngspice} reached the $1600$\,s time limit at $N=512$. The Shockley model also required damping of \texttt{ngspice}'s Newton iterations to converge.

Table~\ref{tab:profiling} reports the median percentage of iteration time spent in each step on the $1024\times1024$ arrays. Iterations account for $71$--$99\%$ of the total runtime of Algorithm~\ref{alg:bis}, with the one-time factorization of $A\Gamma^{-1}A^{\top}$ accounting for the remainder. The steps comprise branch \emph{resolvent evaluation} (line 2), \emph{linear system solution} using the Cholesky factor (line 3), sparse \emph{matrix--vector products} with $A$ and $A^{\top}$ (lines 3 and 4), \emph{convergence checks} for KCL and KVL, and the relaxed \emph{iterate update} (line 4). Resolvent evaluation and the linear system solution account for $62$--$71\%$ of the iteration time, and optimizing these steps as discussed in Remark~\ref{rem:cost} offers substantial speedups.

\begin{table}[t]
  \centering
  \caption{Share (\%) of one iteration of Algorithm~\ref{alg:bis} spent in each step on the $1024\times1024$ arrays, median over 10 random circuits.}
  \label{tab:profiling}
  \footnotesize
\begin{tabular}{@{}lccc@{}}
\toprule
 & 1R & Ideal 1D1R & Shockley 1D1R \\
\midrule
Resolvent evaluation    & 29.5 & 21.3 & 36.4 \\
Linear system solution  & 36.3 & 41.1 & 34.4 \\
Matrix--vector products & 9.9  & 9.9  & 8.1  \\
Convergence checks      & 20.3 & 23.4 & 17.5 \\
Iterate update          & 4.0  & 4.4  & 3.6  \\
\bottomrule
\end{tabular}
\end{table}

\section{Conclusion}\label{sec:conclusion}

The paper has developed an operator splitting method for computing operating points of nonlinear resistive networks. The behavioral formulation separates the branch relations from Kirchhoff's laws and casts circuit simulation as a
zero-finding problem. The resulting relaxed Peaceman--Rachford
iteration uses scattering variables and branch port
resistances, requiring the solution of independent scalar branch equations
and a linear system with a fixed weighted graph Laplacian
whose factorization can be reused. Convergence is established
for maximal monotone branch relations under suitable
relaxation whenever an operating point exists. Contractive
branch reflections yield a linear convergence bound
independent of network topology. Numerical experiments on
crossbar arrays illustrate the computational advantages.

Several directions remain open. Circuits with dynamic elements call for the same analysis in Hilbert spaces of signals. Further work includes inexact solves of the Laplacian system by iterative methods, the selection of the port resistances for nonlinear devices from data, and extensions of our formulation to dynamic optimization over system behaviors~\cite{padoan2026dynamic}.

\bibliographystyle{ieeetr}
\bibliography{references}

@article{Momeni2025,
	title = {Training of physical neural networks},
	volume = {645},
	copyright = {2025 Springer Nature Limited},
	url = {https://www.nature.com/articles/s41586-025-09384-2},
	doi = {10.1038/s41586-025-09384-2},
	number = {8079},
	urldate = {2025-10-27},
	journal = {Nature},
	publisher = {Nature Publishing Group},
	author = {Momeni, Ali and Rahmani, Babak and Scellier, Benjamin and Wright, Logan G. and McMahon, Peter L. and Wanjura, Clara C. and Li, Yuhang and Skalli, Anas and Berloff, Natalia G. and Onodera, Tatsuhiro and Oguz, Ilker and Morichetti, Francesco and del Hougne, Philipp and Le Gallo, Manuel and Sebastian, Abu and Mirhoseini, Azalia and Zhang, Cheng and Markovi\'c, Danijela and Brunner, Daniel and Moser, Christophe and Gigan, Sylvain and Marquardt, Florian and Ozcan, Aydogan and Grollier, Julie and Liu, Andrea J. and Psaltis, Demetri and Alù, Andrea and Fleury, Romain},
	month = sep,
	year = {2025},
	pages = {53--61},
}

@incollection{Liu2019,
  author    = {Liu, Ming and Liu, Qi and Lv, Hangbing and Long, Shibing},
  title     = {Resistive Switching Devices: Mechanism, Performance and Integration},
  booktitle = {Handbook of Memristor Networks},
  editor    = {Chua, Leon and Sirakoulis, Georgios Ch. and Adamatzky, Andrew},
  publisher = {Springer},
  address   = {Cham, Switzerland},
  pages     = {843--911},
  year      = {2019},
  doi       = {10.1007/978-3-319-76375-0_30}
}

@book{golub2013matrix,
  author    = {Golub, Gene H. and Van Loan, Charles F.},
  title     = {Matrix Computations},
  edition   = {4th},
  publisher = {Johns Hopkins Univ. Press},
  address   = {Baltimore, MD, USA},
  year      = {2013}
}

@article{sebastian2020,
  author  = {Sebastian, Abu and Le Gallo, Manuel and Khaddam-Aljameh, Riduan and Eleftheriou, Evangelos},
  title   = {Memory devices and applications for in-memory computing},
  journal = {Nature Nanotechnol.},
  volume  = {15},
  number  = {7},
  pages   = {529--544},
  year    = {2020}
}

@misc{shahhosseini2025splitting,
  author = {Shahhosseini, Amir and Chaffey, Thomas and Sepulchre, Rodolphe},
  title  = {Operator-splitting methods for neuromorphic circuit simulation},
    howpublished = {arXiv:2505.22363},
  year   = {2025}
}

@inproceedings{shahhosseini2025varmetric,
  author    = {Shahhosseini, Amir and Burger, Thiago and Sepulchre, Rodolphe},
  title     = {Variable metric splitting methods for neuromorphic circuit simulation},
  booktitle = {Proc. 64th IEEE Conf. Decis. Control},
  pages     = {968--973},
  year      = {2025}
}

@article{ho1975mna,
  author  = {Ho, Chung-Wen and Ruehli, Albert E. and Brennan, Pierce A.},
  title   = {The modified nodal approach to network analysis},
  journal = {IEEE Trans. Circuits Syst.},
  volume  = {22},
  number  = {6},
  pages   = {504--509},
  year    = {1975}
}

@techreport{nagel1975spice2,
  author      = {Nagel, Laurence W.},
  title       = {{SPICE2}: A computer program to simulate semiconductor circuits},
  institution = {Univ. California},
  address     = {Berkeley, CA, USA},
  number      = {UCB/ERL M520},
  year        = {1975}
}

@book{najm2010circuit,
  author    = {Najm, Farid N.},
  title     = {Circuit Simulation},
  publisher = {Wiley},
  address   = {Hoboken, NJ, USA},
  year      = {2010}
}

@article{davis2010klu,
  author  = {Davis, Timothy A. and Palamadai Natarajan, Ekanathan},
  title   = {Algorithm 907: {KLU}, a direct sparse solver for circuit simulation problems},
  journal = {ACM Trans. Math. Softw.},
  volume  = {37},
  number  = {3},
  pages   = {36:1--36:17},
  year    = {2010}
}

@article{willems2007behavioral,
  author  = {Willems, Jan C.},
  title   = {The behavioral approach to open and interconnected systems},
  journal = {IEEE Control Syst. Mag.},
  volume  = {27},
  number  = {6},
  pages   = {46--99},
  year    = {2007}
}

@article{hachtel1971sparse,
  author  = {Hachtel, Gary D. and Brayton, Robert K. and Gustavson, Fred G.},
  title   = {The sparse tableau approach to network analysis and design},
  journal = {IEEE Trans. Circuit Theory},
  volume  = {18},
  number  = {1},
  pages   = {101--113},
  year    = {1971}
}

@book{chua1987linear,
  author    = {Chua, Leon O. and Desoer, Charles A. and Kuh, Ernest S.},
  title     = {Linear and Nonlinear Circuits},
  publisher = {McGraw-Hill},
  address   = {New York, NY, USA},
  year      = {1987}
}

@book{anderson1973network,
  author    = {Anderson, Brian D. O. and Vongpanitlerd, Sumeth},
  title     = {Network Analysis and Synthesis: A Modern Systems Theory Approach},
  publisher = {Prentice-Hall},
  address   = {Englewood Cliffs, NJ, USA},
  year      = {1973}
}

@book{belevitch1968classical,
  author    = {Belevitch, Vitold},
  title     = {Classical Network Theory},
  publisher = {Holden-Day},
  address   = {San Francisco, CA, USA},
  year      = {1968}
}

@book{bollobas1998modern,
  author    = {Bollob{\'a}s, B{\'e}la},
  title     = {Modern Graph Theory},
  volume    = {184},
  publisher = {Springer},
  address   = {New York, NY, USA},
  year      = {1998}
}

@article{minty1960,
  author  = {Minty, George J.},
  title   = {Monotone networks},
  journal = {Proc. Roy. Soc. London A},
  volume  = {257},
  number  = {1289},
  pages   = {194--212},
  year    = {1960}
}

@article{chaffey2024monotone,
  author  = {Chaffey, Thomas and Sepulchre, Rodolphe},
  title   = {Monotone one-port circuits},
  journal = {IEEE Trans. Autom. Control},
  volume  = {69},
  number  = {2},
  pages   = {783--796},
  year    = {2024}
}

@article{chaffey2023monotoneskew,
  author  = {Chaffey, Thomas and Banert, Sebastian and Giselsson, Pontus and Pates, Richard},
  title   = {Circuit analysis using monotone+skew splitting},
  journal = {Eur. J. Control},
  volume  = {74},
  pages   = {100854},
  year    = {2023}
}

@article{quan2025srg,
  author  = {Quan, Jiajun and Evens, Brecht and Sepulchre, Rodolphe and Patrinos, Panagiotis},
  title   = {Scaled relative graphs for nonmonotone operators with applications in circuit theory},
  journal = {Eur. J. Control},
  volume  = {86},
  pages   = {101335},
  year    = {2025}
}

@article{spingarn1983,
  author  = {Spingarn, Jonathan E.},
  title   = {Partial inverse of a monotone operator},
  journal = {Appl. Math. Optim.},
  volume  = {10},
  number  = {3},
  pages   = {247--265},
  year    = {1983}
}

@article{eckstein1992,
  author  = {Eckstein, Jonathan and Bertsekas, Dimitri P.},
  title   = {On the {Douglas--Rachford} splitting method and the proximal point algorithm for maximal monotone operators},
  journal = {Math. Program.},
  volume  = {55},
  pages   = {293--318},
  year    = {1992}
}

@article{tellegen1952,
  author  = {Tellegen, Bernard D. H.},
  title   = {A general network theorem, with applications},
  journal = {Philips Res. Rep.},
  volume  = {7},
  pages   = {259--269},
  year    = {1952}
}

@article{fettweis1986,
  author  = {Fettweis, Alfred},
  title   = {Wave digital filters: Theory and practice},
  journal = {Proc. IEEE},
  volume  = {74},
  number  = {2},
  pages   = {270--327},
  year    = {1986}
}

@article{bernardini2019,
  author  = {Bernardini, Alberto and Werner, Kurt J. and Smith, Julius O. and Sarti, Augusto},
  title   = {Generalized wave digital filter realizations of arbitrary reciprocal connection networks},
  journal = {IEEE Trans. Circuits Syst. I},
  volume  = {66},
  number  = {2},
  pages   = {694--707},
  year    = {2019}
}

@article{bernardini2025,
  author  = {Bernardini, Alberto and Giampiccolo, Riccardo and Bozzo, Enrico and Fontana, Federico},
  title   = {Wave digital extended fixed-point solvers for circuits with multiple one-port nonlinearities},
  journal = {IEEE Trans. Circuits Syst. I},
  volume  = {72},
  number  = {5},
  pages   = {2314--2326},
  year    = {2025}
}

@book{bauschke2017convex,
  author    = {Bauschke, Heinz H. and Combettes, Patrick L.},
  title     = {Convex Analysis and Monotone Operator Theory in Hilbert Spaces},
  edition   = {2nd},
  publisher = {Springer},
  address   = {Cham, Switzerland},
  year      = {2017}
}

@article{lions1979,
  author  = {Lions, Pierre-Louis and Mercier, Bertrand},
  title   = {Splitting algorithms for the sum of two nonlinear operators},
  journal = {SIAM J. Numer. Anal.},
  volume  = {16},
  number  = {6},
  pages   = {964--979},
  year    = {1979}
}

@book{ryu2022,
  author    = {Ryu, Ernest K. and Yin, Wotao},
  title     = {Large-Scale Convex Optimization: Algorithms and Analyses via Monotone Operators},
  publisher = {Cambridge Univ. Press},
  address   = {Cambridge, U.K.},
  year      = {2022}
}

@article{giselsson2017linear,
  author  = {Giselsson, Pontus and Boyd, Stephen},
  title   = {Linear convergence and metric selection for {Douglas--Rachford} splitting and {ADMM}},
  journal = {IEEE Trans. Autom. Control},
  volume  = {62},
  number  = {2},
  pages   = {532--544},
  year    = {2017}
}

@article{chen2013crossbar,
  author  = {Chen, An},
  title   = {A comprehensive crossbar array model with solutions for line resistance and nonlinear device characteristics},
  journal = {IEEE Trans. Electron Devices},
  volume  = {60},
  number  = {4},
  pages   = {1318--1326},
  year    = {2013}
}

@article{ielmini2018memory,
  author  = {Ielmini, Daniele and Wong, H.-S. Philip},
  title   = {In-memory computing with resistive switching devices},
  journal = {Nature Electron.},
  volume  = {1},
  number  = {6},
  pages   = {333--343},
  year    = {2018}
}

@article{chen2019edge,
  author  = {Chen, Wei-Hao and Dou, Chunmeng and Li, Kai-Xiang
             and Lin, Wei-Yu and Li, Pin-Yi and Huang, Jian-Hao
             and Wang, Jing-Hong and Wei, Wei-Chen and Xue, Cheng-Xin
             and Chiu, Yen-Cheng and King, Ya-Chin and Lin, Chorng-Jung
             and Liu, Ren-Shuo and Hsieh, Chih-Cheng and Tang, Kea-Tiong
             and Yang, J. Joshua and Ho, Mon-Shu and Chang, Meng-Fan},
  title   = {{CMOS}-integrated memristive non-volatile
             computing-in-memory for {AI} edge processors},
  journal = {Nature Electronics},
  volume  = {2},
  number  = {9},
  pages   = {420--428},
  year    = {2019},
  doi     = {10.1038/s41928-019-0288-0}
}

@book{bullo2022lectures,
  author    = {Bullo, Francesco},
  title     = {Lectures on Network Systems},
  edition   = {1.6},
  publisher = {Kindle Direct Publ.},
  year      = {2022}
}

@incollection{davis2016rates,
  author    = {Davis, Damek and Yin, Wotao},
  title     = {Convergence rate analysis of several splitting schemes},
  booktitle = {Splitting Methods in Communication, Imaging, Science, and Engineering},
  editor    = {Glowinski, Roland and Osher, Stanley J. and Yin, Wotao},
  publisher = {Springer},
  address   = {Cham, Switzerland},
  pages     = {115--163},
  year      = {2016}
}

@inproceedings{kyng2016approximate,
  author    = {Kyng, Rasmus and Sachdeva, Sushant},
  title     = {Approximate {G}aussian elimination for {L}aplacians: Fast, sparse, and simple},
  booktitle = {Proc. 57th Annu. IEEE Symp. Found. Comput. Sci.},
  pages     = {573--582},
  year      = {2016}
}

@article{mahey1995proximal,
  author  = {Mahey, Philippe and Oualibouch, Said and Pham Dinh, Tao},
  title   = {Proximal decomposition on the graph of a maximal monotone operator},
  journal = {SIAM J. Optim.},
  volume  = {5},
  number  = {2},
  pages   = {454--466},
  year    = {1995}
}

@book{vanderschaft2017l2gain,
  author    = {van der Schaft, Arjan},
  title     = {{$L_2$}-Gain and Passivity Techniques in Nonlinear Control},
  edition   = {3rd},
  publisher = {Springer},
  address   = {Cham, Switzerland},
  year      = {2017}
}

@article{willems1972dissipative,
  author  = {Willems, Jan C.},
  title   = {Dissipative dynamical systems {P}art {I}: General theory},
  journal = {Arch. Rational Mech. Anal.},
  volume  = {45},
  number  = {5},
  pages   = {321--351},
  year    = {1972}
}

@article{bauschke2012firmly,
  author  = {Bauschke, Heinz H. and Moffat, Sarah M. and Wang, Xianfu},
  title   = {Firmly Nonexpansive Mappings and Maximally Monotone
             Operators: Correspondence and Duality},
  journal = {Set-Valued Var. Anal.},
  volume  = {20},
  number  = {1},
  pages   = {131--153},
  year    = {2012},
  doi     = {10.1007/s11228-011-0187-7}
}

@incollection{rockafellar1995monotone,
  author    = {Rockafellar, R. T.},
  title     = {Monotone relations and network equilibrium},
  booktitle = {Variational Inequalities and Network Equilibrium Problems},
  editor    = {Giannessi, F. and Maugeri, A.},
    publisher = {Plenum Press},
  address   = {New York, NY, USA},
  pages     = {271--288},
  year      = {1995},
  doi       = {10.1007/978-1-4899-1358-6_21}
}

@inproceedings{scellier2024fast,
  author    = {Scellier, B.},
  title     = {A fast algorithm to simulate nonlinear resistive networks},
    booktitle = {Proc. 41st Int. Conf. Mach. Learn.},
  series    = {Proc. Mach. Learn. Res.},
  volume    = {235},
  pages     = {43477--43503},
  year      = {2024},
  url       = {https://proceedings.mlr.press/v235/scellier24a.html}
}

@inproceedings{padoan2025split,
  author    = {Tang, Y. and Cenedese, C. and Rimoldi, A.
               and D{\"o}rfler, F. and Lygeros, J. and Padoan, A.},
  title     = {{Split-as-a-Pro}: behavioral control via operator
               splitting and alternating projections},
  booktitle = {Proc. 23rd Eur. Control Conf.},
  pages     = {1495--1501},
  year      = {2025},
  doi       = {10.23919/ECC65951.2025.11187008}
}

@unpublished{padoan2026dynamic,
  author = {Tang, Y. and Cenedese, C. and Rimoldi, A.
            and D{\"o}rfler, F. and Lygeros, J. and Padoan, A.},
  title  = {Dynamic optimization over system behaviors},
  note   = {In preparation},
  year   = {2026}
}

@article{huang2024memristor,
  author  = {Huang, Y. and Ando, T. and Sebastian, A.
             and Chang, M.-F. and Yang, J. J. and Xia, Q.},
  title   = {Memristor-based hardware accelerators for artificial intelligence},
  journal = {Nat. Rev. Electr. Eng.},
  volume  = {1},
  number  = {5},
  pages   = {286--299},
  year    = {2024},
  doi     = {10.1038/s44287-024-00037-6}
}

@misc{chaffey2025circuit,
  author       = {Thomas Chaffey},
  title        = {Circuit realization and hardware linearization
                  of monotone operator equilibrium networks},
  howpublished = {arXiv:2509.13793},
  year         = {2025},
  doi          = {10.48550/arXiv.2509.13793},
  url          = {https://arxiv.org/abs/2509.13793}
}

@misc{ngspice,
  title        = {{ngspice}: Open Source {SPICE} Simulator},
  howpublished = {\url{https://ngspice.sourceforge.io/}},
  note         = {Accessed: Nov. 2, 2026}
}

\end{document}